\documentclass[a4paper,UKenglish]{lipics-v2021}
\hideLIPIcs
\nolinenumbers
\usepackage{amsmath,amssymb,booktabs,tikz}
\usetikzlibrary{arrows.meta,positioning}
\usepackage{thmtools}
\usepackage{algorithm}
\usepackage{algpseudocode}
\algrenewcommand\algorithmicrequire{\textbf{Input:}}
\algrenewcommand\algorithmicensure{\textbf{Output:}}

\newcommand{\Pref}{\mathsf{Pref}}
\newcommand{\Suff}{\mathsf{Suff}}
\newcommand{\Factor}{\mathsf{Fact}}
\newcommand{\lce}{\operatorname{LCE}}
\newcommand{\lcs}{\operatorname{LCS}}

\newcommand{\LSFL}{\mathsf{LSF}_\ell}
\newcommand{\LSFR}{\mathsf{LSF}_r}

\newcommand{\sfcut}{\mathsf{sf}}

\newcommand{\str}[1]{\texttt{#1}}

\title{Longest Unbordered Factors in Linear Time\thanks{This is an AI-assisted paper.}}
\titlerunning{Longest Unbordered Factors in Linear Time}
\author{Takuya Mieno}{University of Electro-Communications, Japan}{}{}{}
\authorrunning{T. Mieno}
\Copyright{Takuya Mieno}

\begin{document}
\maketitle

\begin{abstract}
  A border of a string is a nonempty proper prefix that is also a suffix.
  A string is unbordered if it has no border.
  The length of a longest unbordered factor is closely related to the smallest period of the string.
  While the smallest period can be computed in linear time,
  the best previously known algorithm for computing a longest unbordered factor takes
  $O(n \log n)$ worst-case time for a string of length $n$.
  Whether this computational problem can be solved in worst-case linear time has remained open.
  In this paper, we resolve this open problem for integer alphabets of size $n^{O(1)}$
  by presenting an algorithm that computes all occurrences of longest unbordered factors in $O(n)$ worst-case time.
  Our algorithm uses a combinatorial property of longest unbordered factors and
  scans the input string in both directions to capture all occurrences achieving the global maximum 
  without computing the local maximum at every starting position.
\end{abstract}

\section{Introduction}\label{sec:introduction}

A \emph{border} of a string $T$ is a nonempty proper prefix of $T$ that is also a suffix of $T$.
A string is said to be \emph{unbordered} if it has no border.
Let $\mu(T)$ denote the length of a longest unbordered factor of $T$.
This quantity is closely related to the \emph{smallest period} of $T$.
For a string $T$ of length $n$, an integer $p$ with $1\le p\le n$ is a \emph{period} of $T$
if $T[i]=T[i+p]$ for all $0\le i<n-p$.
We denote the smallest period of $T$ by $\pi(T)$.
Note that it always holds that $\mu(T)\le\pi(T)$.
The smallest period can be computed in $O(n)$ time using the border array,
which stores the length of the longest border of each prefix.
Periodicity is a central concept in stringology and
has been widely studied in both combinatorics on words and the design of string algorithms.

In 1979, Ehrenfeucht and Silberger~\cite{EhrenfeuchtS79} asked
under what conditions $\mu(T)=\pi(T)$ holds.
They conjectured that $\mu(T)\le n/2$ implies $\mu(T)=\pi(T)$.
Assous and Pouzet~\cite{AssousP79} disproved this conjecture in the same year.
The relation between these two quantities continued to be studied for three decades.
Finally, Holub and Nowotka~\cite{HolubN09,HolubN12} resolved the combinatorial problem by proving that $\mu(T)=\pi(T)$ whenever $\mu(T)\le 3n/7$.
This bound is asymptotically tight.
See~\cite{HolubN12} and references therein for further historical background.

Although $\pi(T)$ can be computed in linear time,
it has remained open whether $\mu(T)$ can also be computed in worst-case linear time.
An $O(n^2)$-time algorithm is straightforward:
the border arrays of all suffixes of $T$ indicate which factors are unbordered.
Loptev et al.~\cite{LoptevKS15} proposed an algorithm with average-case running time $O(n+n^2/\sigma^4)$ for uniformly random strings over an alphabet of size $\sigma$.
Gawrychowski et al.~\cite{GawrychowskiKSS15} proposed an algorithm with $O(n\log n)$ average-case running time and another with $O(n\sqrt{n})$ worst-case running time.
Cording et al.~\cite{CordingGKK21} proposed an algorithm with average-case running time $O(\sqrt{n\log_\sigma n})$ for uniformly random strings over an alphabet of size $\sigma$.
Kociumaka et al.~\cite{kkmp2018} proposed a deterministic $O(n\log n\log^2\log n)$-time algorithm.
Using the prefix-suffix query data structure of Kociumaka et al.~\cite[Theorem~1.4]{KociumakaRRW24},
which supports constant-time queries after deterministic linear-time preprocessing over
polynomially bounded integer alphabets, reduces this bound to $O(n\log n)$.
More recently, Sekizaki and Mieno~\cite{SekizakiM26} studied the problem on run-length encoded strings.

In this paper, we propose a linear-time algorithm that reports
all occurrences of longest unbordered factors.
Our main result is the following.

\begin{restatable}{theorem}{maintheorem}\label{thm:main}
  Given a string $T$ of length $n$ over an integer alphabet of size $n^{O(1)}$,
  we can compute all occurrences of longest unbordered factors of $T$ in $O(n)$ time.
\end{restatable}

Our algorithm builds on the right-to-left scan of Kociumaka et al.~\cite{kkmp2018}.
Their algorithm computes the LUF array,
which stores the length of the Longest Unbordered Factor starting at each position.
That is, their algorithm computes the local maximum at each starting position.
Our algorithm only needs to find all occurrences achieving the global maximum $\mu(T)$.
Thus, the candidate at each position need not be locally optimal.
This weaker requirement allows us to use simple updates that take constant time per position.
Our scans use the LSF arrays of~\cite{kkmp2018} and consider an additional candidate ending at a \emph{square-free cut}, defined in Section~\ref{sec:preliminaries}.
The key combinatorial property is stated in Lemma~\ref{lem:unique} for uniquely occurring longest unbordered factors.
For factors occurring more than once, we use Duval's periodicity result~(Lemma~\ref{lem:duval}).
Together, these properties show that two scans,
one on $T$ and one on its reversal, capture all occurrences attaining the global maximum.
The scans are easy to implement\footnote{The preprocessing requires suffix sorting and run enumeration, for which existing implementations can be used.}.

Section~2 introduces the notation and basic properties of unbordered strings.
Section~3 presents our algorithm and an example.
Section~4 proves its correctness.
Section~5 shows the linear-time preprocessing.

\section{Preliminaries}\label{sec:preliminaries}
\subsection{Notation and computation model}
For non-negative integers $i, j$ with $i \le j$, we define $[i.. j]=\{i, i+1, \ldots, j\}$ as the set of consecutive integers from $i$ to $j$. Let $[i.. j) = [i.. j]\setminus \{j\}$.
We call a set $[i.. j]$ an interval.
In what follows, all numerical variables denote integers.

Let $\Sigma = [0.. \sigma)$ be an \emph{integer alphabet} of size $\sigma$.
An element of $\Sigma$ is called a \emph{character}.
An element of $\Sigma^\star$ is called a \emph{string}.
The \emph{length} of a string $T$ is denoted by $|T|$, i.e., $T \in \Sigma^{|T|}$ for any string $T$.
The concatenation of strings is denoted by juxtaposition.
If $T = XYZ$ holds for strings $T,X,Y,Z\in\Sigma^\star$,
then $X, Y$, and $Z$ are called a \emph{prefix}, a \emph{factor}, and a \emph{suffix} of $T$, respectively.
Each is said to be \emph{proper} if its length is smaller than $|T|$.
The first character of $T$ is the $0$-th character.
For a string $T$ and integers $i, j \in [0.. |T|)$, we denote by $T[i]$ the $i$-th character of $T$ and
by $T[i.. j]$ the factor of $T$ that starts at $i$ and ends at $j$.
For convenience, let $T[i.. j]$ be the empty string if $i > j$.
Further, let $T[i.. j)$ denote $T[i.. j-1]$.
Let $\Pref(T) = \{T[0.. j] \mid j \in [0.. |T|)\}$ denote the set of nonempty prefixes of a string $T$.
Similarly, let $\Suff(T)$ and $\Factor(T)$ denote the sets of nonempty suffixes and factors of a string $T$, respectively.
The \emph{reversal} of a string $T$ is $T[|T|-1]T[|T|-2]\cdots T[0]$ and is denoted by $T^R$.
In what follows, we arbitrarily fix a nonempty string $T$ of length $n \ge 2$.

A nonempty interval $[s.. e] \subseteq [0.. n)$ is called an \emph{occurrence} of a string $P$ in $T$
if $P = T[s.. e]$ holds.
Then, positions $s$ and $e$ are called the starting and the ending positions of the occurrence, respectively.
A nonempty proper prefix $B$ of $T$ is called a \emph{border} of $T$ if $B$ is also a suffix of $T$.
We say that $T$ has a border $B$ if $B$ is a border of $T$.
A nonempty string is \emph{unbordered} if the string has no border.
Note that any string of length one is unbordered.
We denote by $\mu(T)$ the length of a longest unbordered factor of $T$.
A positive integer $p \in [1.. n]$ is called a \emph{period} of a string $T$ if
$T[i] = T[i+p]$ holds for all $i \in [0.. n-p)$.
We say that $T$ has a period $p$ if $p$ is a period of $T$.
Let $\pi(T)$ denote the smallest period of $T$.
It is well known that $T$ has a border of length $b$ iff $T$ has a period $n-b$.
Thus, $T$ is unbordered iff $\pi(T) = n$.

A \emph{cut} of $T$ is an integer $q\in[0..n]$.
It denotes the boundary between $T[0..q)$ and $T[q..n)$.
A nonempty string of the form $UU$ is called a \emph{square}.
A cut $q$ of $T$ is called the \emph{center} of an occurrence of a square $UU$ if $T[q-|U|.. q) = T[q.. q+|U|) = U$ holds.
A cut $q$ of $T$ is said to be square-free if no square has center $q$.
In other words, $q$ is a square-free cut iff $T[q-r.. q) \ne T[q.. q+r)$ for all $1\le r \le \min\{q, n-q\}$.
For cuts $i, j \in [0.. n]$,
let $\lce(i,j)$ be the length of the longest common prefix of $T[i.. n)$ and $T[j.. n)$, and
let $\lcs(i,j)$ be the length of the longest common suffix of $T[0.. i)$ and $T[0.. j)$.
Nonempty strings $X, Y$ are said to be mutually \emph{nonintersecting} if
$\Pref(X) \cap \Suff(Y) = \emptyset$,
$\Suff(X) \cap \Pref(Y) = \emptyset$,
$X \not\in \Factor(Y)$, and
$Y \not\in \Factor(X)$.
Equivalently, $X$ and $Y$ are mutually nonintersecting if and only if
no string contains intersecting occurrences of $X$ and $Y$, where intersection includes containment.

This paper assumes that the alphabet size $\sigma$ is $n^{O(1)}$,
so we can sort the characters from the input string $T$
and can reduce the alphabet to integers $[0.. n)$ in $O(n)$ time
by using a radix sort.
The model of computation in this paper is a standard word RAM model of word size $\Omega(\log n)$.

\subsection{Basic properties of unbordered strings}\label{sec:basic}

In this subsection, we list some known properties of unbordered strings used later.
The first claim of the next lemma is stated in~\cite[Lemma~3.3]{HarjuN07} without a proof;
we give a proof for completeness.
\begin{lemma}\label{lem:short-extension}
  Let $X$ be a nonempty and unbordered string.
  If $S$ is a nonempty proper suffix of $X$, then $XS$ is unbordered.
  If a character $c$ differs from $X[0]$, then $Xc$ is unbordered.
\end{lemma}
\begin{proof}
  By definition, $|S| < |X|$ holds.
  For the sake of contradiction, we assume that $XS$ has a border $B$.
  If $|B| \le |S|$, $B$ is a suffix of $S$, which is a proper suffix of $X$.
  Also, $B$ is a proper prefix of $X$ since $|B| \le |S| < |X|$.
  Thus $B$ is a border of $X$, a contradiction.
  If $|B| > |S|$, the length-$(|B|-|S|)$ prefix $B'$ of $B$ is a proper suffix of $X$.
  Also, $B'$ is a proper prefix of $X$ since $|B|-|S| < |XS|-|S| = |X|$.
  Thus $B[0.. |B|-|S|)$ is a border of $X$, a contradiction.

  For the string $Xc$, there is no length-$1$ border since $c \ne X[0]$.
  A longer border gives a border of $X$ after deleting its final character, a contradiction.
  Hence, $Xc$ is unbordered.
\end{proof}

It is known that any string can be decomposed into
unbordered prefixes of the string:
\begin{lemma}[Unbordered-decomposition~\cite{duval1982}]\label{lem:decompose}
  For any nonempty string $S$,
  there exists a unique sequence $(U_1, \ldots, U_t)$ of unbordered prefixes of $S$
  such that $S = U_1\cdots U_t$.
  Furthermore, 
  $U_1$ is the longest unbordered prefix of $S$, and
  $U_j$ is an unbordered prefix of $U_1$ for all $j$ with $1 \le j \le t$.
\end{lemma}
We call each $U_j$ in Lemma~\ref{lem:decompose} a \emph{phrase} of the unbordered-decomposition of string $S$.

The following lemma follows from the proof of Lemma 3.6 of Harju and Nowotka~\cite{HarjuN07}.
We give a direct proof for completeness.

\begin{lemma}\label{lem:insertion}
  Let $A,B$ be mutually nonintersecting nonempty strings.
  If $X$ is nonempty and unbordered, then $AXB$ is unbordered or has period $|X|$.
\end{lemma}
\begin{proof}
  Let $x = |X|$, $a=|A|$, $b=|B|$, and $m=a+x+b$.
  Suppose $AXB$ has a border of length $\ell$.
  If $\ell \le a$, the length-$\ell$ prefix $A'$ of $A$ occurs as a suffix of $AXB$,
  which implies that $A'$ is a suffix of $B$ or $B$ is a suffix of $A'$.
  This contradicts that $A$ and $B$ are mutually nonintersecting.
  If $a < \ell < a + b$, then the occurrence of $A$ starting at $m-\ell$ overlaps
  the rightmost $B$, a contradiction.
  Thus $\ell \ge a+b$.
  The border of length $\ell \ge a+b$ gives a period $p = m-\ell \le x$.
  It is also a period of $X$ that is a factor of $AXB$.
  Since $X$ is unbordered, $p = x$ holds.
\end{proof}

The following lemma was shown by Duval~\cite{duval1982}.
\begin{lemma}[Corollary 2.13 of \cite{duval1982}]\label{lem:duval}
  If a longest unbordered factor $X$ of $T$ occurs at least twice in $T$,
  then it holds that $\pi(T)=|X|=\mu(T)$.
\end{lemma}

\section{The algorithm}\label{sec:algorithm}

In our algorithm, we use three arrays:
the longest successor factor array $\LSFL$,
LSF-Reference array $\LSFR$, and square-free cut array $\sfcut$.
The first two arrays are defined in~\cite{kkmp2018}.
For $i\in[0.. n)$, define $\LSFL[i] = \max \{\lce(i,j)\mid i<j\le n\}$.
When $\LSFL[i] > 0$, let $\LSFR[i]$ be the largest $j > i$ with $\lce(i, j)=\LSFL[i]$.
We call $\LSFR[i]$ the \emph{rightmost reference} of $i$.
Note that the LSF-Reference array is used only when $\LSFL[i] > 0$.
Furthermore, for each cut $q \in [0.. n]$, we set $\sfcut[q] = 1$ if $q$ is a square-free cut of $T$, and $\sfcut[q]=0$ otherwise.
In particular, $\sfcut[0] = \sfcut[n] = 1$.
All arrays and lengths of common extensions refer to the current string,
either $T$ or $T^R$.

Our main procedure is shown in Algorithm~\ref{alg:main}.
By running the subroutine $\textsc{Scan}(T)$,
we obtain the candidate array $C_0$ that stores
the lengths of unbordered factors of $T$ starting at every position.
A symmetric procedure $\textsc{Scan}(T^R)$ gives another candidate array $C_1$ for the reversal of $T$.
Then, the longest candidates within $C_0$ and $C_1$ are actually the longest unbordered factors of $T$.
The correctness will be proven later.
\begin{algorithm}[h!]
  \caption{Main procedure.}
  \label{alg:main}
  \begin{algorithmic}[1]
    \Require String $T$.
    \Ensure All occurrences of longest unbordered factors of $T$.
    \State $C_0 \gets \Call{Scan}{T}$
    \State $C_1 \gets \Call{Scan}{T^R}$
    \State $\mathit{maxlen} \gets \max\{\max C_0,\max C_1\}$
    \State $\mathit{occ} \gets$ all distinct intervals in $T$ of length $\mathit{maxlen}$ represented by $C_0$ or $C_1$
    \State \Return $\mathit{occ}$
  \end{algorithmic}
\end{algorithm}
\begin{algorithm}[t]
  \caption{Function $\textsc{Scan}(T)$: preprocessing and right-to-left scan over $T$.}
  \label{alg:scan}
  \begin{algorithmic}[1]
    \Require String $T$ of length $n$.
    \Ensure Integer array $C$ described in the main text.
    \State Compute arrays $\LSFL, \LSFR$, and $\sfcut$ for $T$.
    \Comment{Preprocessing (cf. Section~\ref{sec:preprocessing}).}
    \State Create an integer array $C$ of length $n$.
    \For{$i = n-1,n-2,\ldots,0$}
    \If{$\LSFL[i] = 0$}
    \State $C[i] \gets n-i$
    \Else
    \State $r \gets \LSFR[i]$
    \If{$C[r] \le \LSFL[i]$}
    \State $C[i] \gets C[r]$
    \Comment{Copy the candidate.}
    \Else
    \State $C[i] \gets C[r] + (r-i)$
    \Comment{Extend the candidate.}
    \EndIf
    \If{$\sfcut[r] = 1$}
    \State $C[i] \gets \max\{C[i],r-i\}$
    \Comment{Consider the cut candidate.}
    \EndIf
    \EndIf
    \EndFor
    \State \Return $C$
  \end{algorithmic}
\end{algorithm}

In the subroutine $\textsc{Scan}(T)$ shown in Algorithm~\ref{alg:scan},
we first compute arrays $\LSFL, \LSFR$, and $\sfcut$ as preprocessing (Line~1).
At position $i$, we compute the length $C[i]$ of one unbordered factor that starts at $i$.
We do \emph{not} require this factor to be longest among those starting at $i$.
We process positions from right to left.
If $\LSFL[i]=0$, the first character of $T[i..n)$ occurs only once in this suffix.
Thus, the suffix is unbordered, and we set $C[i] = n-i$ (Line~5).
Otherwise, let $r=\LSFR[i]$.
Since $r > i$, the candidate length $C[r]$ has already been computed.
If $C[r] \le \LSFL[i]$, the unbordered string $T[r.. r+C[r])$ also occurs at $i$, and we copy its length (Line~9).
Otherwise, we extend the candidate to start at the current position $i$.
Its new length is $C[r] + (r - i)$ (Line~11).
As shown in Lemma~\ref{lem:sound},
a border of this longer string would also border the candidate at $r$.
Furthermore, we add one more candidate when $r$ is a square-free cut of $T$ (Lines~13--15).
Its interval is $[i.. r)$ and its length is $r - i$.
Again, as shown in Lemma~\ref{lem:sound},
if this candidate has a border, it creates a square centered at $r$,
which leads to a contradiction.
We keep the longer one of the available candidates.

In Section~\ref{sec:correctness}, we show that every stored candidate is unbordered and that every occurrence of a longest unbordered factor is obtained by at least one of the two scans for $T$ and $T^R$.

Excluding preprocessing at Line~1 of Algorithm~\ref{alg:scan}, the algorithm clearly takes $O(n)$ time.
This $O(n)$ time bound includes mapping occurrences in $T^R$ back to $T$ and removing duplicates at Line~4 of Algorithm~\ref{alg:main}.
Section~\ref{sec:preprocessing} shows that the preprocessing also takes $O(n)$ time,
so the total running time is $O(n)$.

\paragraph*{Example}

Consider $T=\str{abaabbaa}$ of length $8$.
The arrays for $T$ are shown below.
A dash means that the reference is unused.
\[
  \begin{array}{c|rrrrrrrrr}
    i&0&1&2&3&4&5&6&7&8\\\hline
    T[i]&\str{a}&\str{b}&\str{a}&\str{a}&\str{b}&\str{b}&\str{a}&\str{a}&\\
    \LSFL[i]&2&3&2&1&1&0&1&0&\\
    \LSFR[i]&3&5&6&7&5&\text{--}&7&\text{--}&\\
    \sfcut[i]&1&1&1&0&1&0&1&0&1\\
    C_0[i]&1&3&4&1&4&3&1&1&
  \end{array}
\]

Let us see the details for positions $5$ to $1$.
At position $5$, we set $C_0[5] \gets 8-5 = 3$ because $\LSFL[5]=0$~(Line~5).
At position $4$, Line~11 extends this candidate and we get $C_0[4] \gets C_0[\LSFR[4]]+(\LSFR[4]-4)= C_0[5] + (5-4) = 4$.
At position $3$, it just copies $C_0[3] \gets C_0[7] = 1$.
At position $2$, it first copies $C_0[2] \gets C_0[\LSFR[2]] = C_0[6] = 1$ at Line~9.
Since $\sfcut[6] = 1$, it then raises $C_0[2]$ to $6 - 2 = 4$ at Line~14.
At position $1$, it just copies $C_0[1] \gets C_0[5]=3$.
The maximum value in $C_0$ for $T$ is $4$.
The reverse string is $T^R=\str{aabbaaba}$,
with candidate array $C_1=(4,2,6,3,3,2,2,1)$.
The entry $C_1[2]=6$ yields $(T^R[2.. 7])^R = T[0.. 5]=\str{abaabb}$.
This factor is unbordered and has maximum length $\mu(T)=6$.

\section{Correctness}\label{sec:correctness}
This section gives a proof of the correctness of Algorithm~\ref{alg:main}.
We first establish a useful property of the endpoints of longest unbordered factors.
\begin{lemma}\label{lem:endpoints}
  Let $X$ be a longest unbordered factor of $T$.
  The ending position of the rightmost occurrence of $X$ plus one is a square-free cut of $T$.
  Symmetrically, the starting position of the leftmost occurrence of $X$ is also a square-free cut of $T$.
\end{lemma}
\begin{proof}
  We show the former claim for the rightmost endpoint.
  The latter one follows by reversal.
  Let $[i.. q)$ be the rightmost occurrence with $T[i.. q) = X$.
  For the sake of contradiction, we assume $T[q-|U|.. q+|U|) = UU$ for some nonempty string $U$.
  If $|U| < |X|$, then $U$ is a proper suffix of $X$ and $XU$ occurs at position $i$.
  By Lemma~\ref{lem:short-extension}, $XU$ is unbordered, which contradicts the maximality of $X$.
  If $|U| \ge |X|$, then $X$ is a suffix of $U$.
  The square $UU$ implies the occurrence of $X$ starting at position $q+|U|-|X| > i$,
  which contradicts that $[i.. q)$ is the rightmost occurrence of $X$.
\end{proof}

The next lemma states that the candidate values obtained by Algorithm~\ref{alg:scan} are valid.
\begin{lemma}\label{lem:sound}
  For every $i \in [0.. n)$, the factor $T[i.. i+C[i])$ is unbordered.
\end{lemma}
\begin{proof}
  Since $C[i]$ is updated only in the iteration for $i$ in Algorithm~\ref{alg:scan},
  it suffices to show that every iteration updates $C[i]$ so that $T[i.. i+C[i])$ is unbordered.
  We use induction on $i$ in decreasing order.
  The base case $i = n-1$ is obvious since the length-$1$ suffix $T[n-1]$ is unbordered.

  We say that an update $C[i] \gets x$ is \emph{valid} if $T[i.. i+x)$ is unbordered.
  \begin{description}
    \item[Update at Line 5.] If $\LSFL[i] = 0$, then $T[i]$ occurs only once in $T[i.. n)$, so this suffix is unbordered. Thus $C[i] \gets n-i$ is valid.
    \item[Update at Line 9.] If $\LSFL[i] > 0$, let $r = \LSFR[i]$ and 
      let $U = T[r.. r+C[r])$ be the candidate string starting at $r$.
      By the induction hypothesis, string $U$ is unbordered.
      If $C[r] \le \LSFL[i]$, string $U$ also occurs at $i$.
      Namely, $T[i.. i+C[r])$ is unbordered.
      Thus $C[i] \gets C[r]$ is valid.
    \item[Update at Line~11.] In addition, if $C[r] > \LSFL[i]$, assume that $T[i.. r+C[r])$ has a border $B$ of length $b$.
      Then $b \le \LSFL[i] < C[r]$ holds since the occurrence of $B$ as a suffix of $T[i..r+C[r])$ starts after $i$.
      Since $b\le\LSFL[i]$ and $r=\LSFR[i]$, we have $B=T[i..i+b)=T[r..r+b)$.
      Thus, $B$ is a proper prefix of $U$ because $b<C[r]$. 
      Further, $B$ is a suffix of $T[i..r+C[r])$ and thus of $U$, i.e., $U$ is bordered, a contradiction.
      Therefore, $T[i.. r+C[r])$ is unbordered.
      Thus $C[i] \gets C[r]+(r-i)$ is valid.
    \item[Update at Line~14.] Finally, if $\sfcut[r] = 1$, assume that $T[i.. r)$ has a border $B'$ of length $b'$, which again satisfies $b' \le \LSFL[i]$.
      It follows that $T[r-b'..r)=B'=T[i..i+b')=T[r..r+b')$,
      which gives a square centered at $r$, a contradiction.
      Therefore, $T[i..r)$ is unbordered.
      Both the current candidate and $T[i..r)$ are unbordered,
      so the update $C[i]\gets\max\{C[i],r-i\}$ is valid.
  \end{description}
  Thus, $T[i..i+C[i])$ is unbordered after the iteration for $i$, completing the induction.
\end{proof}

The next lemma gives a sufficient condition for the scan to find the longest unbordered factor starting at a given position.
\begin{lemma}\label{lem:propagation}
  Let $T[i.. q)$ be nonempty and unbordered, with $\sfcut[q] = 1$.
  If $q = n$ or $\LSFL[q] < \lce(i, q)$, then $C[i] \ge q-i$ holds.
\end{lemma}
\begin{proof}
  We prove the claim by induction on $i$ in decreasing order,
  simultaneously for all cuts $q$ satisfying the assumptions.
  For the base case $i=n-1$, the assumptions imply $q=n$,
  and hence $C[n-1]=1=q-i$.

  If $\LSFL[i] = 0$, then $C[i] = n-i \ge q-i$~(Line~5 of Algorithm~\ref{alg:scan}).
  Otherwise, let $r=\LSFR[i]$.
  The rightmost choice of $r$ gives $\LSFL[r] < \LSFL[i]$.

  We next show that $r \le q$.
  If $q = n$, then $r < q$ holds automatically.
  Thus, we consider the other case where $q < n$ and $\LSFL[q] < \lce(i,q)$.
  Let $k = \lce(i,q)$.
  Then $k \le \LSFL[i]$ by definition.
  Assume $r > q$~(see also Figure~\ref{fig:propagation}). 
  Since $r = \LSFR[i]$, the suffixes starting at $i$ and $r$ share a prefix of length $\LSFL[i] \ge k$.
  The suffixes starting at $i$ and $q$ also share a prefix of length $k$ by the definition of $k=\lce(i,q)$.
  Thus, the length-$k$ prefixes starting at $q$ and $r$ are equal.
  Since $r > q$, this gives a later occurrence of the length-$k$ prefix of $T[q..n)$, so $\LSFL[q] \ge k$.
  This contradicts the assumption $\LSFL[q] < k$.
  Thus we have $r \le q$.
  \begin{figure}[h]
    \centering
    \includegraphics[width=0.8\linewidth]{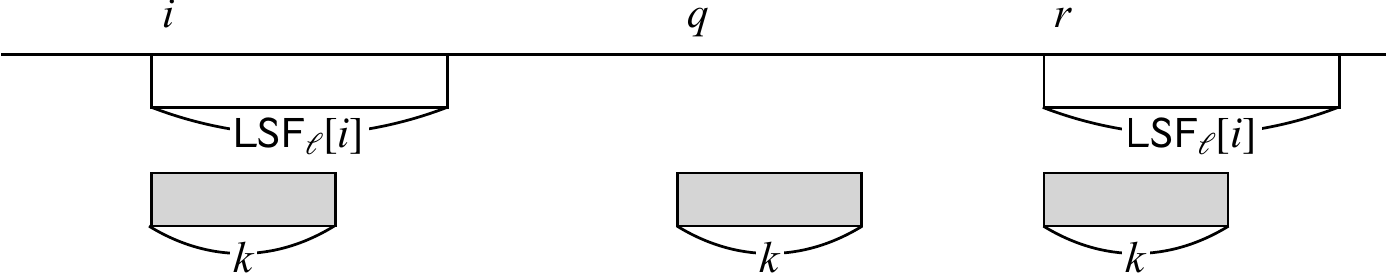}
    \caption{Illustration of the case $q < r$.
    The gray blocks represent the same string of length $k$.} 
    \label{fig:propagation}
  \end{figure}

  If $r = q$, then $C[i] \ge r-i = q-i$ holds due to Line~14 of Algorithm~\ref{alg:scan}.

  The remaining case to be considered is $r < q$.
  Assume that $T[r.. q)$ has a border $B$ of length $b$.
  If $b \le \LSFL[i]$, then $B$ is also a border of $T[i.. q)$, a contradiction.
  Otherwise, $b > \LSFL[i]$, and the two occurrences of $B$ give $b \le \LSFL[r]$,
  contradicting $\LSFL[r]<\LSFL[i]$.
  Thus, $T[r..q)$ is unbordered.
  Recall that $\sfcut[q]=1$ holds by assumption.
  If $q<n$, again let $k = \lce(i,q)$.
  Since $i < q$, the definition of $\LSFL[i]$ gives $k \le \LSFL[i]$.
  By the definition of $r$, we have $\LSFL[i] = \lce(i,r)$.
  Thus, the length-$k$ prefixes starting at $r$ and $q$ are both equal to $T[i..i+k)$.
  Hence $k = \lce(i,q) \le \lce(r, q)$ holds.
  The assumption $\LSFL[q] < \lce(i,q)$ gives $\LSFL[q] < \lce(r,q)$. 
  Hence the induction hypothesis at $r$ gives $C[r] \ge q-r$.
  If we assume $\LSFL[i] \ge q-r$, then $T[r..q)$ is a border of $T[i..q)$, a contradiction.
  Thus, $\LSFL[i] < q-r$ holds.
  Therefore, $C[r] > \LSFL[i]$ holds,
  so the extension rule at Line~11 is applied and yields $C[i]\ge C[r]+(r-i)\ge q-i$.
\end{proof}

Together with Lemma~\ref{lem:endpoints}, the next lemma shows that every uniquely occurring longest unbordered factor of $T$ satisfies the conditions of Lemma~\ref{lem:propagation} in at least one of the two directions.

\begin{lemma}[Key lemma]\label{lem:unique}
  Let $T[i.. q) $ be a longest unbordered factor that occurs only once in $T$, with $0 < i < q < n$.
  Let $\LSFL'$ be the longest successor factor array of $T^R$.
  Then $\LSFL[q] < \lce(i, q)$ or $\LSFL'[n-i] < \lcs(i, q)$.
\end{lemma}
\begin{proof}
  Let $X = T[i.. q)$, $m=|X|$, $k=\lce(i,q)$, and $h=\lcs(i,q)$.
  Since $T[i.. q)$ is the longest unbordered factor, $T[i] = T[q]$ holds by Lemma~\ref{lem:short-extension}.
  Thus, $k$ is positive. Similarly, $h$ is also positive.
  In addition, uniqueness of $T[i.. q)$ gives $h, k < m$, since otherwise $X$ occurs again starting at position $q$ or $i-m$.
  In what follows, we assume on the contrary that $k \le \LSFL[q]$ and $h\le \LSFL'[n-i]$. 

  Let $Y = X[0.. k)$ and $Z = X[m-h.. m)$.
  Since $\LSFL[q] \ge k$, there are at least two occurrences of $Y$ in $T[q.. n)$.
  Let $R$ be the shortest prefix of $T[q.. n)$ containing two occurrences of $Y$.
  Note that $R$ starts and ends with $Y$ and contains $Y$ exactly twice.
  Similarly, let $L$ be the shortest suffix of $T[0.. i)$ containing two occurrences of $Z$.
  Also, $L$ starts and ends with $Z$ and contains $Z$ exactly twice.
  Then, string $LXR$ is a factor of $T$, with $|L| > h$ and $|R| > k$.
  The lengths of the longest matches give
  $R[k] \ne X[k]$ and $L[|L|-h-1] \ne X[m-h-1]$.
  Since $|X| = \mu(T)$, $X$ is the longest unbordered prefix of $XR$ and
  is the longest unbordered suffix of $LX$.
  Now, we show the next claim:
  \begin{claim}\label{claim:outer}
    Every unbordered prefix of $R$ has length at most $k$,
    and every unbordered suffix of $L$ has length at most $h$.
  \end{claim}
  \begin{proof}[Proof of Claim~\ref{claim:outer}]
    By Lemma~\ref{lem:decompose}, the unbordered-decomposition of $XR$ starts with $X$.
    Removing this first phrase gives $R=U_1\cdots U_t$,
    where each $U_j$ is an unbordered prefix of $X$. 
    Assume that $U_j$ is longer than $Y$.
    Then, $U_j$ starts with $Y$.
    Since $Y$ occurs exactly twice in $R$,
    the starting position of $U_j$ is either $0$ or $|R|-k$ in $R$.
    The latter case is impossible because $|U_j| > k$.
    Hence, $U_j$ starts at the first position of $R$.
    However, such an occurrence of $U_j$ gives $R[k] = U_j[k] = X[k]$, a contradiction.
    Thus, every phrase has length at most $k$.

    Consider any prefix $V$ of $R$ with $|V| > k$.
    We now prove that $V$ has a border.
    Let $U_\star$ be the phrase of the above decomposition of $R$ that contains the last position of $V$.
    Let $H$ be the nonempty prefix of the phrase $U_\star$ ending at the last position of $V$~(see also Figure~\ref{fig:VhasBorder}).
    \begin{figure}[tb]
      \centering
      \includegraphics[width=0.8\linewidth]{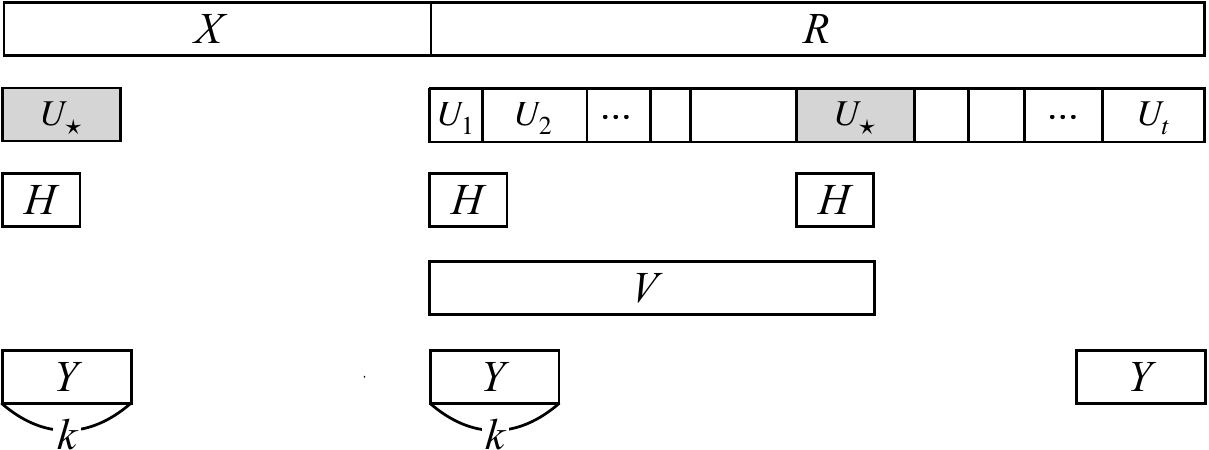}
      \caption{A prefix $V$ of $R$ with $|V|>k$ and its border $H$.} 
      \label{fig:VhasBorder}
    \end{figure}    
    Then $H$ is a suffix of $V$.
    Further, $H$ is a prefix of $X$ since $H$ is a prefix of $U_\star$, which is a prefix of $X$.
    Since $1 \leq |H| \le |U_\star| \le  k < |V|$, $H$ is a prefix of $Y$,
    which is a prefix of $V$.
    Therefore, $H$ is a border of $V$, i.e., $V$ is bordered.

    Applying the same argument to $X^R$ and $L^R$, with $h$ in place of $k$,
    shows that every unbordered suffix of $L$ has length at most $h$.
  \end{proof}

  Let us consider the string $LR$. 
  Suppose position $|L|$ is not a square-free cut of $LR$.
  Then, let $UU$ be the shortest square centered at position $|L|$ in $LR$.
  Its root $U$ is unbordered, since any border of $U$ implies a shorter square at the same center.
  By Claim~\ref{claim:outer}, we have $|U| \le \min\{h, k\} < m$.
  Since $L$ ends with $Z$ of length $h$ and $R$ starts with $Y$ of length $k$,
  the string $U$ is a border of $X$, a contradiction.
  Thus, position $|L|$ is a square-free cut of $LR$.
  Now, we show the next claim:
  \begin{claim}\label{claim}
    There are a nonempty suffix $A$ of $L$ and a nonempty prefix $B$ of $R$
    that are mutually nonintersecting and satisfy $|A| > h$ or $|B| > k$.
  \end{claim}
  \begin{proof}[Proof of Claim~\ref{claim}.]
    Recall that $L$ and $R$ have borders of lengths $h > 0$ and $k > 0$, respectively.
    Since $|L|$ is a square-free cut of $LR$,
    the last character of $L$ and the first character of $R$ differ.
    Thus, there exists a pair of a nonempty suffix of $L$ and a nonempty prefix of $R$
    that are mutually nonintersecting. 
    Among all such pairs, choose a suffix $A$ and a prefix $B$ with maximum total length $|A|+|B|$.
    Let $a=|A|$, $b=|B|$.
    To prove that these $A, B$ satisfy the condition of the statement,
    we assume on the contrary that $a \le h$ and $b \le k$.

    The border of $L$ ensures multiple occurrences of $A$ in $L$.
    Let $U$ be the suffix of $L$ starting with the second rightmost occurrence of $A$ in $L$.
    By definition, $A$ occurs exactly twice in $U$.
    We prove that $B$ must occur in $U$:
    If $B$ does not occur in $U$, then $U$ and $B$ are mutually nonintersecting:
    any occurrence of $B$ intersecting $U$ either lies inside $U$ or intersects its leftmost or rightmost occurrence of $A$ in $U$.
    The former case contradicts the absence of $B$ in $U$;
    the latter case contradicts the nonintersection of $A$ and $B$.
    Therefore, the pair $U, B$ has greater total length than $A, B$, a contradiction.
    Thus $B$ occurs in $U$.

    Let $P$ be the suffix of $U$ starting with the rightmost occurrence of $B$ in $U$.
    Since $A$ and $B$ are mutually nonintersecting, every occurrence of $B$ in $U$ lies between its leftmost and rightmost occurrences of $A$, disjoint from both.
    Thus, $P$ excludes the leftmost occurrence of $A$ in $U$ and retains the rightmost one.
    By the choice of the rightmost occurrence of $B$ in $U$, string $P$ contains exactly one occurrence of each of $A$ and $B$.

    Symmetrically, let $V$ be the prefix of $R$ ending with the second leftmost occurrence of $B$ in $R$.
    By the symmetric argument, $A$ occurs in $V$.
    Let $Q$ be the prefix of $V$ ending with its leftmost occurrence of $A$.
    This occurrence lies between the two occurrences of $B$, disjoint from both.
    Hence, $Q$ retains the leftmost occurrence of $B$ in $V$ and excludes the rightmost one,
    and contains exactly one occurrence of each of $A$ and $B$.
    Both $P$ and $Q$ start with $B$ and end with $A$, so $|P|,|Q|\ge a+b$.
    See Figure~\ref{fig:claim} for illustration.
    \begin{figure}[tb]
      \centering
      \includegraphics[width=0.8\linewidth]{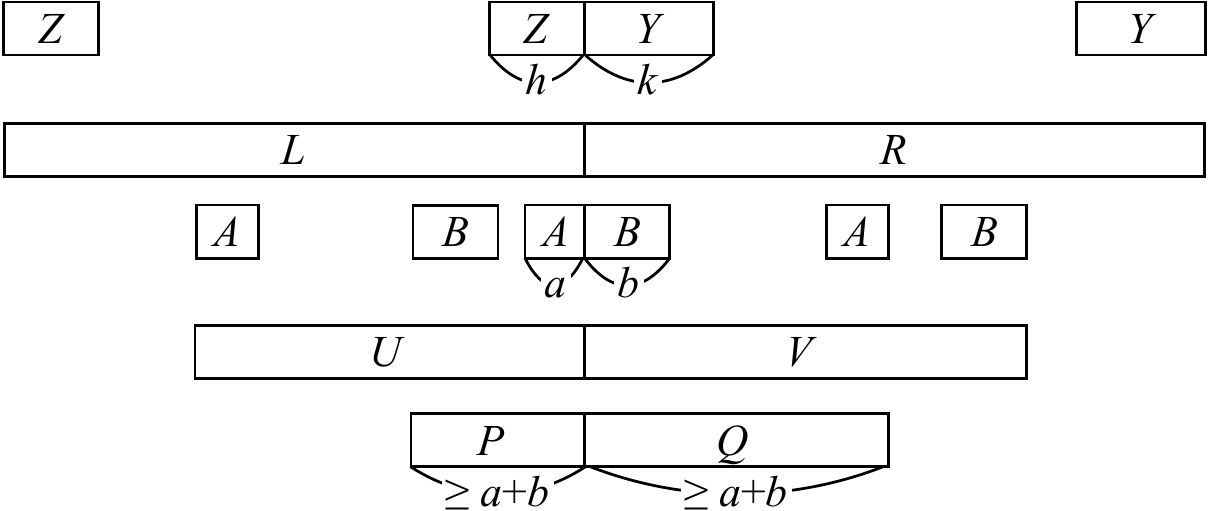}
      \caption{
      Construction of $U,V,P,Q$ in $LR$ under the assumptions $a\le h$ and $b\le k$.} 
      \label{fig:claim}
    \end{figure}

    If $P=Q$, then $PQ$ forms a square centered at $|L|$ in $LR$, a contradiction.
    Otherwise, suppose an occurrence of $P$ intersects an occurrence of $Q$ in some string.
    If their starting positions coincide, the shorter occurrence is contained in the longer one.
    Their distinct ending positions give two occurrences of $A$ in the longer one, a contradiction.
    If their starting positions differ, the leftmost occurrence of $B$ in the later one starts inside the earlier one.
    If this occurrence of $B$ lies entirely inside the earlier occurrence,
    it gives a second occurrence of $B$ there, a contradiction.
    Otherwise, it crosses the right boundary of the earlier occurrence and
    intersects its rightmost occurrence of $A$, contradicting the nonintersection of $A$ and $B$.
    Thus, $P$ and $Q$ are mutually nonintersecting.
    Since $P$ is a suffix of $L$ and $Q$ is a prefix of $R$, and $|P|+|Q|\ge 2(a+b)>a+b$,
    this contradicts the maximality of the pair $A,B$.
    Therefore, $a>h$ or $b>k$. 
  \end{proof}
  Let $A,B$ satisfy the conditions of Claim~\ref{claim}.
  The factor $AXB$ is longer than $m = \mu(T)$, so it is bordered.
  By Lemma~\ref{lem:insertion}, $AXB$ has period $m = |X|$. 
  Since $m = q-i$, this periodicity gives
  $T[i-|A|.. i) = T[q-|A|.. q)$ and $T[i.. i+|B|) = T[q.. q+|B|)$.
  Thus, $|A| \le \lcs(i, q) = h$ and $|B| \le \lce(i, q) = k$,
  contradicting the condition that $|A|>h$ or $|B|>k$.

  Therefore, $\LSFL[q] < k$ or $\LSFL'[n-i] < h$. 
  This concludes the proof of Lemma~\ref{lem:unique}.
\end{proof}

We are ready to show the completeness of the two-direction scans.
\begin{lemma}\label{lem:correctness}
  Let $m = \mu(T)$ be the longest unbordered factor length of $T$
  and let $C_0$ and $C_1$ be the arrays computed in Algorithm~\ref{alg:main}.
  For every $i \in [0.. n-m]$, the factor $T[i.. i+m)$ is unbordered if and only if
  $C_0[i] = m$ or $C_1[n-i-m] = m$.
\end{lemma}
\begin{proof}
  ($\Leftarrow$)
  Lemma~\ref{lem:sound} straightforwardly gives that
  $C_0[i] = m$ or $C_1[n-i-m] = m$ implies $T[i.. i+m)$ is unbordered.

  ($\Rightarrow$) 
  Fix any longest unbordered string $X$ occurring in $T$.
  We show that every occurrence of $X$ is represented by a candidate of length $m$ in $C_0$ or $C_1$.
  We consider two cases based on the uniqueness of $X$ in $T$.
  \begin{enumerate}
    \item Suppose $X$ occurs only once in $T$.
      Let $T[i.. q) = X$ be the occurrence of $X$.
      By Lemma~\ref{lem:endpoints}, both endpoints of $T[i.. q)$ are square-free cuts of $T$,
      i.e., $\sfcut[i] = \sfcut[q] = 1$.
      If $q = n$ or $i = 0$,
      we can apply Lemma~\ref{lem:propagation} to $X$ in $T$ or $X^R$ in $T^R$.
      Otherwise, Lemma~\ref{lem:unique}, together with $\sfcut[i] = \sfcut[q] = 1$,
      allows us to apply Lemma~\ref{lem:propagation} to $X$ in $T$ or to $X^R$ in $T^R$. 
      Thus, applying Lemma~\ref{lem:propagation} in one of the two directions,
      we have $C_0[i] \ge q-i = m$ or $C_1[n-i-m] = C_1[n-q] \ge q-i = m$.
      By Lemma~\ref{lem:sound}, both candidate lengths are at most $m$.
      Hence $C_0[i] = m$ or $C_1[n-q] = m$.
    \item Suppose $X$ occurs at least twice in $T$.
      In this case, we show that the right-to-left scan on $T$ alone captures every occurrence of $X$. 
      By Lemma~\ref{lem:duval}, the smallest period of $T$ is $\pi(T)=m$.
      Firstly, let us consider the rightmost occurrence of $X$ in $T$, say $T[s.. q)=X$.
      Its right endpoint $q$ is square-free by Lemma~\ref{lem:endpoints}.
      Due to the periodicity, $n - q < m$ holds, since otherwise it gives another occurrence of $X$ at $q$.
      If $q = n$, again, we can directly apply Lemma~\ref{lem:propagation}.
      Otherwise, periodicity gives $\lce(s, q) = n - q$,
      whereas $\LSFL[q] < n - q = |T[q.. n)|$.
      In either case, applying Lemma~\ref{lem:propagation} to $X$ in $T$
      gives $C_0[s] \ge m$, and hence $C_0[s] = m$.

      We now prove that $C_0[i]=m$ holds for all earlier occurrences of $X$,
      by induction on their starting positions in decreasing order.
      At such a position $i < s$, a later occurrence of $X$ gives $\LSFL[i] \ge m$.
      Thus $r := \LSFR[i]$ satisfies $\lce(i,r) = \LSFL[i] \ge m$,
      so $X$ also occurs at position $r>i$.
      By induction, $C_0[r] = m \le \LSFL[i]$.
      Line~9 of Algorithm~\ref{alg:scan} assigns $C_0[r]$ to $C_0[i]$, giving $C_0[i] = m$.
      Consequently, the right-to-left scan on $T$ captures
      every occurrence of $X$.
  \end{enumerate} 
  In both cases, every occurrence of $X$ is captured by at least one of the two scans.
\end{proof}

Lemma~\ref{lem:correctness} ensures that at least one candidate from $C_0, C_1$ has length $m = \mu(T)$.
Together with Lemma~\ref{lem:sound}, this gives $\mathit{maxlen}=\mu(T)$ for the value computed in Algorithm~\ref{alg:main}.

\section{Linear-time preprocessing}\label{sec:preprocessing}
This section shows $O(n)$-time algorithms to compute the three arrays
$\LSFL, \LSFR$, and $\sfcut$ for a given string $T$ of length $n \ge 2$.
Recall that the alphabet is assumed to be a set of integers $[0..n)$.
Kociumaka et al.~\cite{kkmp2018} showed that
arrays $\LSFL$ and $\LSFR$ can be computed in $O(n)$ time
with simple modifications to the construction algorithm for
the longest previous factor array given in~\cite{CrochemoreI08}.

It remains to compute the square-free cut array $\sfcut$.
The idea is to represent the set of square centers as the union of $O(n)$ intervals obtained from \emph{runs}~(a.k.a.~\emph{maximal repetitions}).
Then, the count at $q$ is positive if and only if $\sfcut[q] = 0$.
A run is a factor $T[i.. j)$ of $T$ with the smallest period $p \le (j-i)/2$ such that
(1) $i = 0$ or $\pi(T[i-1.. j)) > p$ and
(2) $j = n$ or $\pi(T[i.. j+1)) > p$.
We denote by $([i.. j), p)$ a run $T[i.. j)$ with the smallest period $p$.
It is known that there are at most $n$ runs in a string of length $n$ and
all runs can be enumerated in $O(n)$ time~\cite{BannaiIINTT17}.
In addition, each run $([i.. j), p)$ gives a square centered at every cut $q \in [i+p.. j-p]$,
because $T[q-p.. q) = T[q.. q+p)$ holds.
Conversely, for any square centered at $q$ that has the smallest period $p$,
there is exactly one run $([i.. j), p)$ with the same smallest period $p$
such that the square is a factor of $T[i.. j)$ and thus $q \in [i+p.. j-p]$.
Thus, the union of the intervals $[i+p..j-p]$ over all runs
$([i.. j), p)$ is exactly the set of square centers in $T$.
We compute this union with a difference array $\delta[0.. n+1)$, initially zero.
For each run $([i.. j), p)$, we increment the value $\delta[i+p]$ at the left endpoint of $[i+p.. j-p]$ and
decrement the value $\delta[j-p+1]$ immediately after the right endpoint of $[i+p.. j-p]$.
The prefix sum $\sum_{t=0}^{q}\delta[t]$ counts the intervals containing cut $q$.
For each $q\in[0..n]$, we set $\sfcut[q]=0$ if $\sum_{t=0}^{q}\delta[t]>0$,
and $\sfcut[q]=1$ otherwise.
This procedure takes $O(n)$ time since
each run contributes exactly two updates,
and all prefix sums take $O(n)$ time in total.

Therefore, the preprocessing can be done in $O(n)$ time,
and thus Algorithm~\ref{alg:main} runs in $O(n)$ time.
Together with Lemma~\ref{lem:correctness}, we have proven our main theorem restated below:
\maintheorem*

\section*{AI Disclosure}

This manuscript was prepared with AI assistance.
The author's main contribution was to improve the readability of the paper. 
An initial version of the algorithm was obtained from a six-hour run of ChatGPT dots on 9 October 2026.
ChatGPT 6 Astra (Ultra) was then used to organize the material into a draft paper.
The author reviewed the generated draft, simplified the algorithm,
checked the technical arguments in detail, investigated the relationship to previous work,
created the figures, and substantially restructured the paper.
The author takes full responsibility for the content of this paper.

\end{document}